# Solid-state heating using the multicaloric effect in multiferroics

Melvin M. Vopson[1,*], Yuri K. Fetisov[2], Ian Hepburn[3]

[1] Faculty of Technology, School of Mathematics and Physics, University of Portsmouth, Portsmouth, UK
[2] MIREA - Russian Technological University, Moscow, Russia
[3] Department of Space & Climate Physics, Faculty of Mathematics & Physical Sciences, University College London, UK

* Correspondence: melvin.vopson@port.ac.uk;

**Abstract:** The multicaloric effect is defined as the adiabatic temperature change in multiferroic materials induced by the application of an external electric or magnetic field, and it was first proposed in 2012. The multicaloric effects in multiferroics, as well as other similar caloric effects in single-ferroics, have been the focus of much research due to their potential commercialization in solid-state refrigeration. In this short communication article we examine the thermodynamics of the multicaloric effect for solid-state heating applications. A possible thermodynamic multicaloric heating cycle is proposed, and then implemented to estimate the solid-state heating effect for a known electrocaloric system. This work offers a path to the implementation of caloric and multicaloric effects to efficient heating systems and we offer a theoretical estimate of the upper limit of the temperature change achievable in a multicaloric cooling or heating effect.

## 1. Introduction

Solid-state caloric effects [1-6] manifest as a temperature change within a given physical system in response to adiabatic changes of internal or external variables such as: volume, strain, magnetization or polarization. These temperature changes can be either heating or cooling, depending on the sequence of the applied excitation, i.e. application or removal of the specific control parameter.

There is in fact a particular interest in the solid-state magnetocaloric or electrocaloric cooling because they offer the prospect of vibration free, low noise, efficient and environmentally friendly refrigeration, including room temperature refrigeration applications but also ultra-low cryogenics [7]. The ability to develop solid state cooling technologies is also attractive for integrating cooling devices into electronic and micro-electronic components [8].

Driven by their huge commercialization potential, the research in solid-state caloric effects has accelerated with most of the efforts concentrated on finding suitable materials that display large temperature changes, leading to the report of giant caloric effects [1,9,10]. The research efforts on discovering huge caloric effects in single–ferroic materials, have been aided by parallel research in a class of materials used for caloric effects called multiferroics [11]. Multiferroic materials display simultaneously multiple ferroic order phases. Since each ferroic order phase can facilitate a wide range of applications, it is normally expected that materials displaying multiple combined order phases, would offer enhanced capabilities leading to an abundance of applications including advanced sensors [12-14], memories [15,16], magnetic recording readers [17,18], transformers [19] and even energy harvesting devices [20]. Some of the most promising applications of multiferroic materials have been detailed in this review article [21], with one of them being its application to solid-state caloric effects. This stimulated the concept of multi-caloric effect in multiferroics, first proposed theoretically in 2012 [22] and followed by a number of other studies revolving around the same concept [23 - 32]. In reference [23] the generalized theory of the giant caloric effects was introduced, which allows one to describe all possible caloric and multicaloric effects including those induced by more exotic means of excitation such as mechanical stress [33, 34]. For historical accuracy, it is important to mention that the term "multicaloric" appeared in the public domain the same year, in an earlier article [35]. While Fahler et al. [35] proposed the possibility of an enhanced caloric effect via magneto-electric coupling in multiferroics, the theoretical justification was presented by Vopson [22] a few months later, as also credited by other authors [36].

The original introduction of the multicaloric effect in multiferroics was done theoretically using a thermodynamic approach. Assuming a multiferroic solid containing electrically and magnetically ordered phases, the differential Gibbs free energy is: $dG = -S \cdot dT - M \cdot dH - P \cdot dE$, where S is the entropy, M is the magnetization and P the polarization of the system. Assuming that this system displays a linear magneto-electric effect characteristic to multiferroics and mathematically defined by the $\alpha$ coupling coefficient $(\partial M/\partial E)_{T,H} = (\partial P/\partial H)_{T,E} = \alpha$, then the electrically (1) and magnetically (2) induced multicaloric effects in a given multiferroic system are expressed as [22]:

$$\Delta T_E = -\frac{T}{C} \cdot \int_{E_i}^{E_f} \left[ \frac{\alpha_e}{\mu_0 \chi^m} \cdot \left(\frac{\partial M}{\partial T}\right)_{H,E} + \left(\frac{\partial P}{\partial T}\right)_{H,E} \right] \cdot dE \quad (1.1)$$

$$\Delta T_H = -\frac{T}{C} \cdot \int_{H_i}^{H_f} \left[ \left(\frac{\partial M}{\partial T}\right)_{H,E} + \frac{\alpha_m}{\varepsilon_0 \chi^e} \cdot \left(\frac{\partial P}{\partial T}\right)_{H,E} \right] \cdot dH \quad (1.2)$$

where $\mu_0$ is the magnetic permeability of vacuum, $\varepsilon_0$ is the dielectric permittivity of vacuum, C is the specific heat capacity of the system per unit volume as $C = T \cdot (\partial S/\partial T)_{H,E}$, and $\chi^m$ and $\chi^e$ are the susceptibilities of the magnetic and polar phase, respectively. The full derivation of relations (1) is given in references [22,23].

Equations (1) indicate that a single excitation field could induce a full multicaloric effect. If the electric field is used to induce the effect, then only equation (1.1) applies, while if a magnetic field is driving the caloric process, then only equation (1.2) applies. Hence, a single excitation delivers a multiple caloric response, i.e. multicaloric. This is very different to ordinary ferroic systems where a single excitation induces a single caloric response.

Since M and P of most ferroic and multiferroic materials decrease with the temperature, i.e. $\partial M/\partial T < 0$ and $\partial P/\partial T < 0$, at constant applied fields, then the multicaloric temperature change given by relations (1) can be positive ($\Delta T_{E,H} > 0$) for adiabatic polarization / magnetization, or negative ($\Delta T_{E,H} < 0$) for adiabatic depolarization / demagnetization, respectively. A closer examination of relations (1) indicates that they are very similar to those describing the electrocaloric and magnetocaloric effects, except that the multicaloric effects contain additional terms, $(\alpha_e/(\mu_0 \cdot \chi^m)) \cdot (\partial M/\partial T)$ and $(\alpha_m/(\varepsilon_0 \cdot \chi^e)) \cdot (\partial P/\partial T)$. These additional terms result from the magneto-electric coupling in multiferroics and they can further enhance the thermal effect, especially in strongly coupled multiferroics. For large multicaloric coupling terms we expect a significant increase in the total temperature change. In fact, according to (1), $\Delta T$ can increase indefinitely with increasing the $\alpha$ coupling coefficient. This, of course, is not possible so the following question arises:

*What is the maximum predicted temperature change in the multicaloric effect?*

In order to answer this question, it is useful to rewrite equations (1) in a further simplified manner by expressing the derivatives of the magneto-electric induced M and P with respect to T as: $(\partial M/\partial T) = \alpha_e \cdot (\varepsilon_0 \cdot \chi^e)^{-1} \cdot (\partial P/\partial T)$ and $(\partial P/\partial T) = \alpha_m \cdot (\mu_0 \cdot \chi^m)^{-1} \cdot (\partial M/\partial T)$. Using these expressions, we obtained the following simplified relations of the electrically and magnetically induced multicaloric effects:

$$\Delta T_E = -\frac{T}{C} \cdot \left( \frac{\alpha_e^2}{\mu_0 \varepsilon_0 \chi^m \chi^e} + 1 \right) \cdot \left( \frac{\partial P}{\partial T} \right)_{H,E} \cdot \Delta E \quad (2.1)$$

$$\Delta T_H = -\frac{T}{C} \cdot \left( \frac{\alpha_m^2}{\mu_0 \varepsilon_0 \chi^m \chi^e} + 1 \right) \cdot \left( \frac{\partial M}{\partial T} \right)_{H,E} \cdot \Delta H \quad (2.2)$$

where $\Delta E = E_f - E_i$ and $\Delta H = H_f - H_i$. Equations (2) show clearly the enhancement of the electrocaloric and magnetocaloric effects in the case of multicaloric effect in multiferroics, with the additional contribution to $\Delta T$ given by the magneto-electric caloric coupling term $(\alpha^2/(\mu_0 \cdot \varepsilon_0 \cdot \chi^m \chi^e) + 1)$. Indeed, in the particular case of a material that does not display any magneto-electric coupling or it is not multiferroic ($\alpha_e = \alpha_m = 0$), equations (2) simply describe the electrocaloric and magnetocaloric effects.

However, the magneto-electric coupling coefficient has a fundamental thermodynamic limit, which is given by the relation $\alpha^2 \leq (\mu_0 \cdot \varepsilon_0 \cdot \chi^m \chi^e)$ [11]. A more recent derivation of this limit was given here [37]. This implies that the magneto-electric caloric coupling term takes always fractional values, $\alpha^2/(\mu_0 \cdot \varepsilon_0 \cdot \chi^m \chi^e) \leq 1$. Imposing this condition in equations (2), an upper limit can be established for the maximum $\Delta T$ expected for the electrically or magnetically induced multicaloric effect, which is twice the temperature change expected for the equivalent electrocaloric or magnetocaloric effects induced by the same excitation fields, if no magneto-electric coupling exists. This offers a significant enhancement of the temperature change in a multicaloric effect relative to single-caloric effects.

The main focus of all previous studies of single-caloric effects in single-ferroic systems has been the development of solid-state cooling technologies. Similarly, the recently proposed multicaloric effect in multiferroics is regarded as the most promising candidate for the development of efficient solid-state refrigeration systems [22,23], far superior to single-caloric effects.

In this short communication article, we take a different approach and, instead of solid state cooling, we propose the application of the multicaloric effect in multiferroics to solid-state heating technologies. The proposed operation principle is explained in figure 1. For clarity we show the cooling diagram (Fig. 1.a) and thermodynamic cycle (Fig. 1.b) (adapted from previously published work [22]), and the newly proposed heating operation principle (Fig. 1.c.d).

We first explain briefly the multicaloric cooling cycle. The system is initially at the operation temperature $T_A$. The system is also assumed initially in thermal equilibrium. Since the operation temperature $T_A$ is selected so that the cooling agent is in a para-ferroic state, then at this state magnetic and electric dipole moments are thermally activated and undergo random fluctuations in a para-multiferroic state (Fig. 1.a). Imposing adiabatic state, a single excitation field H or E is applied to the multiferroic system. The effect of the field application is to align the magnetic and electric dipole moments. This is essentially a transition from disorder (high entropy state) to ordered multiferroic (low entropy state) (transition A-B, Fig. 1.b). Hence, the decrease in the entropy of the system under adiabatic conditions will increase the overall temperature of the system to $T_B = T_A + \Delta T$. This additional temperature could be reduced back to the initial temperature via a heat sink. In this process, the applied E or H field is maintained constant, preventing the magnetic and electric dipoles from reabsorbing heat. The required operating temperature usually dictates the nature of the heat sink, and it is usually a fluid coolant like water for room temperature operation, or a cryogenic liquid for cryogenic cooling. The transition B-C in Fig. 1.b corresponds to the system returning to the initial equilibrium temperature $T_A$ of the heat sink. Using a thermal switch to break the contact with the heat sink, the system returns to adiabatic conditions and the total entropy remains constant again. Simultaneously, the applied H or E field is switched off, corresponding to transition C-D, Fig. 1.b). The field removal initiates an adiabatic demagnetization and depolarization process, causing the magnetic and electric moments to absorb heat as they relax back to equilibrium. Since entropy increases again, the adiabatic condition is fulfilled by decreasing the temperature of the refrigerant to a value lower than the temperature of the heat sink, i.e. $T_D = T_A - \Delta T$. The transition D-A in Fig. 1.b corresponds to the multiferroic refrigerant being placed in thermal contact with the environment being refrigerated, ending the cooling cycle. The solid-state cooling technology and its thermodynamic cycle are well established and essentially applied in identical manner to all the solid-state caloric effects, with the only difference being the caloric material itself and the corresponding excitation force / field.

Our proposed solid-state heating thermodynamic cycle is very similar to the cooling cycle except that it has three stages instead of four (see Figure 1.c, d).

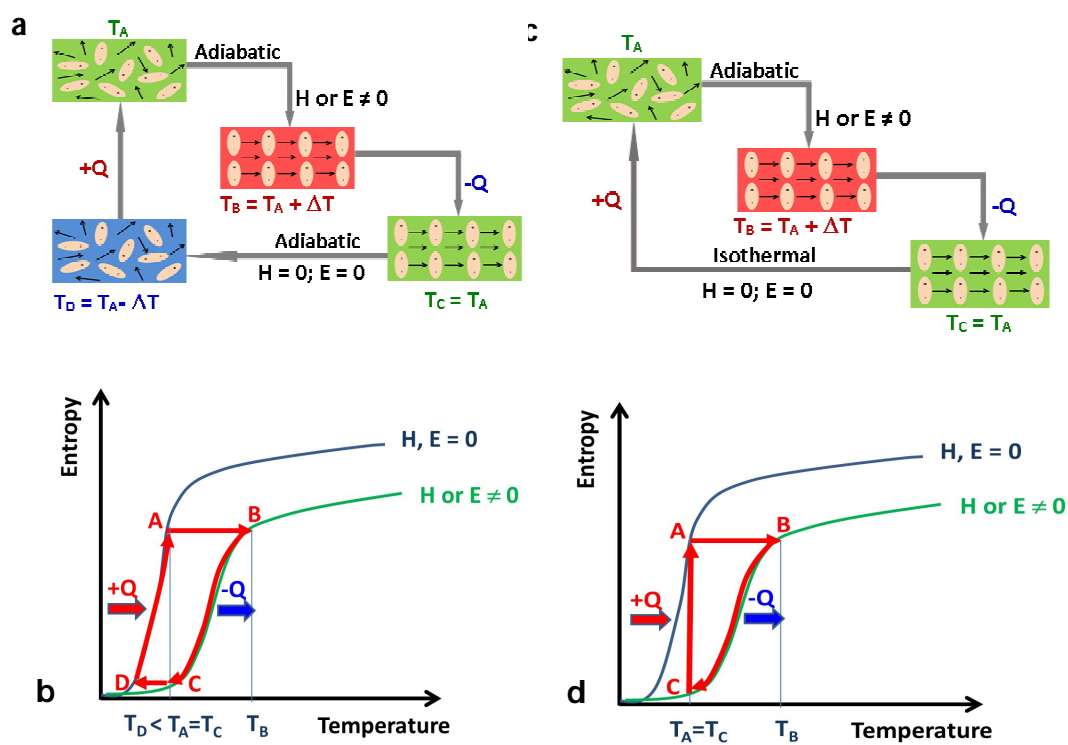

**Figure 1.** Diagrams of the proposed cooling (a, b adapted from ref. [22]), and heating cycles of the solid-state multicaloric effect. a) Schematic of the four stages cooling cycle (A-D) showing the entropy change due to relaxation of the magnetic moments (black arrows) and electric dipoles (ovals); b) The corresponding Brayton cooling cycle; c) Schematic of the three stages heating cycle (A-C); d) The corresponding thermodynamic heating cycle.

The multicaloric heating cycle begins again with the system at the initial operation temperature $T_A$ (Fig. 1.c,d). Upon the application of a H or E field, forcing the magnetic spins and electric dipole moments to align and reducing the entropy of the system, the multiferroic's temperature increases to $T_B = T_A + \Delta T$ (transition A-B, Fig. 1.c,d). While the applied excitation field is still on, the excess temperature is transferred to the environment via a heat sink. For room temperature heating applications, the heat exchange / sink is typically water circulated in contact with the multiferroic heating element. In this process, the system returns to the initial equilibrium temperature $T_A$ given by the heat sink (transition B-C, Fig. 1.c,d). Maintaining thermal contact with the heat sink, the applied H or E field is switched off (transition C-A, Fig. 1.c,d). This is, in effect, an isothermal demagnetization and depolarization process, which causes the spins and electric dipoles to exchange heat with the environment, at constant temperature $T_A$. The multiferroic heating element is then subjected to another field application and the whole cycle is repeated again.

The temperature change ΔT (increase for heating or decrease for cooling), is governed by the set of multicaloric effect equations (1) or (2), also applicable for non multiferroic materials.

By selecting an active multiferroic material displaying order phase transition temperatures of the magnetic and electric phases at room temperature [38,39] (i.e. $T_c^m \approx T_c^e \approx$ 300K), then $\partial M/\partial T, \partial P/\partial T$ and the total entropy change are largest at around 300K. This property combined with a large enough magneto-electric coupling coefficient, would result in significant ΔT changes. In terms of room temperature heating applications for domestic use, this is interesting as it suggests that a domestic heating system operating on the proposed multicaloric heating principle, would only require a temperature change of around ΔT = 10K, in order to ensure that it maintains a constant working temperature of the environment ideal for habitation. Assuming that active multiferroic elements displaying large magneto-electric coupling effects at room temperature are developed, then heating systems operating on this principle would certainly become a reality.

To highlight the potential heating application we have investigated an idealized heating system comprising the electrocaloric material $PbSc_{0.5}Ta_{0.5}O_3$ (PST). Using the data presented in Nair et. al [40] we have constructed an idealized model to investigate the potential output heating power, electrical input power and the resulting coefficient of performance (CoP) for a system operating between 10 °C and 60 °C, these being the respective vales for $T_A$ and $T_B$ of Figure 1.d. As it is typical of currently known electrocaloric materials the temperature change on the application of an electric field is low, for PST the maximum is ~ 4 K at 305 K with 15.8 $V\mu m^{-1}$ electric field [40]. In order to increase the temperature differential between points $T_A$ and $T_B$ (Figure 1.d) we have considered a 13 PST multi-layer system. The resulting thermal output power and electrical input power are presented in Figure 2 as a function of the speed in which the thermodynamic cycle in Figure 1.d can be achieved. The resulting coefficient of performance, i.e. the difference in output power to input power is ~3, giving an efficiency of ~300%. This highlights the high efficiency / CoP due to the ability to extract heat from the low temperature end of the system in the same way that heat pumps extract heat from their surroundings.

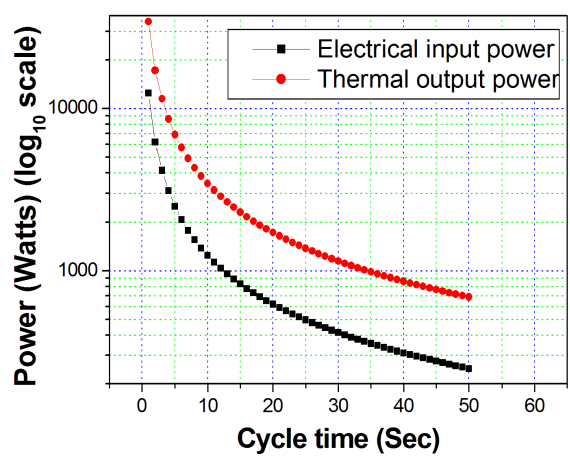

**Figure 2.** Thermal output power and electrical input power for our idealized electrocaloric heater.

A cycle time (path A-B-C in Figure 1.d) of the order of seconds is not unrealistic considering the sub-minute magnetocaloric cycle times presented in Bartlett et. al [41] and the relative ease in which an electric field can be generated compared to a magnetic field. When the same heating principle is applied to a suitable multiferroic active element, up to twice the temperature change is expected, leading to even more efficient heating systems.

The purpose of this communication article is to reemphasize the fundamentals of solid-state refrigeration based on the new concept of multicaloric effect in multiferroics, and to propose a new possible application of this effect to solid state heating, in addition to the refrigeration. Theoretically, both heating and cooling offer the promise of achieving ultra-efficient temperature changes with predicted cooling / heating rates per cycle of up to double the values achieved in electrocaloric or magnetocaloric materials. We therefore hope that this work will stimulate experimental and commercial interest in developing not only solid state cooling based on the multicaloric effect, but also solid state heating technologies.

**Author Contributions:** All authors have contributed equally to preparing this manuscript. All authors have read and agreed to the published version of the manuscript.

**Funding:** MV and IH received no external funding for this research. Research of YF at RTU MIREA was supported by the Russian Science Foundation, project No. 17-12-01435-P.

**Acknowledgments:** MV acknowledges the support received from SMAP, University of Portsmouth to undertake this research.

**Conflicts of Interest:** The authors declare no conflict of interest.

## References

1. L. Manosa, D. Gonzalez-Alonso, A. Planes, E. Bonnot, M. Barrio, J.-L. Tamarit, S. Aksoy, M. Acet, Giant solid-state barocaloric effect in the Ni-Mn-In magnetic shape-memory alloy, Nature Materials **9**, 478 (2010).
2. E. Bonnot, R. Romero, L. Man osa, E. Vives, A. Planes, Elastocaloric Effect Associated with the Martensitic Transition in Shape-Memory Alloys, Phys. Rev. Lett.,**100**, 125901 (2008).
3. K. A. Gschneidner, V. K. Pecharsky, A. O. Tsokol, Recent developments in magnetocaloric materials, Rep. Prog. Phys. **68**, 1479-1539 (2005).
4. J.F. Scott, Electrocaloric Materials, Annu. Rev. Mater. Res. **41**, 1 – 12 (2011).
5. T. Castan, A. Planes, A. Saxena, Phys. Rev. B 85, 144429 (2012).
6. M.S. Reis, Solid State Communications, vol. 152, 921–923 (2012).
7. J. Bartlett, G. Hardy, I.D. Hepburn, C. Brockley-Blatt, P. Coker, E. Crofts, B. Winter, S. Milward, R. Stafford-Allen, M. Brownhill, J. Reed, M. Linde, N. Rando, Improved performance of an engineering model cryogen free double adiabatic demagnetization refrigerator, Cryogenics 50 (2010) 582–590.
8. A. Shakouri, Y. Zhang, On-chip solid-state cooling for integrated circuits using thin-film microrefrigerators, IEEE Transactions on Components and Packaging Technologies, 28 (1) (2005).
9. K.V. Pecharsky, K.A. Gschneider, Giant magnetocaloric effect in $Gd_5(Si_2Ge_2)$ Phys. Rev. Lett. **78**, 4494 - 4497 (1997).
10. A. S. Mischenko, Q. Zhang, J. F. Scott, R. W. Whatmore, N. D.Mathur, Gian Electrocaloric Effect in Thin-Film $PbZr_{0.95}Ti_{0.05}O_3$, Science **311**, 1270- 1271 (2006).
11. M. Fiebig, Revival of the magnetoelectric effect, J. Phys. D: Appl. Phys. **38,** (2005) R123.
12. Gao J, Shen L, Wang Y, Gray D, Li J, Viehland D. Enhanced sensitivity to direct current magnetic field changes in Metglas/$Pb(Mg_{1/3}Nb_{2/3})O_3$–$PbTiO_3$ laminates. J. Appl. Phys. **109**, 074507 (2011).
13. M. Vopsaroiu M.G. Cain, G. Sreenivasulu, G. Srinivasan, A.M. Balbashov, Multiferroic composite for combined detection of static and alternating magnetic fields,   Materials Letters **66,** 282–284 (2012).
14. Gajek M, Bibes M, Fusil S, Bouzehouane K, Fontcuberta J, Barthelemy A, et al. Tunnel junctions with multiferroic barriers. Nature Materials, **6**,   296–302 (2007).
15. Manuel Bibes, Agnès Barthélémy, Multiferroics: Towards a magnetoelectric memory, Nature Materials **7**, 425 - 426 (2008).
16. J. F. Scott, Data storage: Multiferroic memories, Nature Materials 6, 256 - 257 (2007).
17. Vopsaroiu M, Blackburn J, Piniella A, Cain MG. Multiferroic magnetic recording read head technology for 1 Tb/in2 and beyond. J. Appl. Phys., **103**,   07F506 (2008).

18. Yi Zhang, Li Zheng, Deng Chaoyong, Ma Jing, Lin Yuanhua, Ce-Wen Nan. Demonstration of magnetoelectric read head of multiferroic heterostructures. Appl. Phys. Lett., **92**, 152510 (2008).
19. Dong SX, Li JF, Viehland D. Voltage gain effect in a ring-type magnetoelectric laminate. Appl. Phys. Lett. **84**, 4188 (2004).
20. Srivastava V, Song Y, Bhatti K, James RD. The direct conversion of heat to electricity using multiferroic alloys Adv. Energy Mater. **1**, 97–104 (2011).
21. M.M. Vopson, Fundamentals of Multiferroic Materials and Their Possible Applications, Crit. Rev. Solid State Mater. Sci. 40(4), 223–250 (2015).
22. M.M. Vopson, The multicaloric effect in multiferroic materials, Solid State Communications 152, 2067–2070 (2012).
23. M.M. Vopson, Theory of giant-caloric effects in multiferroic materials, J. Phys. D: Appl. Phys. 46 (2013) 345304.
24. H Meng, B Li, W Ren, Z Zhang, Physics Letters A, Volume 377, Issue 7, 567–571 (2013).
25. S. Alpay, J. Mantese, S. Trolier-McKinstry, Q. Zhang, R. W. Whatmore, MRS Bulletin 39, no. 12 (2014) 1099-1111.
26. M.M. Vopson, D. Zhou, G. Caruntu, Multicaloric effect in bi-layer multiferroic composites, Applied Physics Letters, 107(18), p.182905 (2015).
27. A. Planes, T. Castan, A. Saxena, Philosophical Magazine 94, no. 17 (2014) 1893-1908.
28. Y. Liu, W. Jie, J. Pierre-Eymeric, I.C. Infante, J. Kreisel, X. Lou, B. Dkhil, Physical Review B 90, no. 10 (2014): 104107.
29. S. Patel, A. Chauhan, R. Vaish, Applied Physics Letters 107, no. 4 (2015): 042902.
30. M.M. Vopson, The induced magnetic and electric fields' paradox leading to multicaloric effects in multiferroics, Solid State Communications, 231, pp.14-16 (2016).
31. Zhao, YQ., Cao, HX. Multicaloric effect in multiferroic $EuTiO_3$ thin films. J Mater Sci **55,** 5705–5714 (2020).
32. Vopson, M.M., Multicaloric effect: An outlook, Physica B: Condensd Matter 2017, 513, 103-105.
33. P. O. Castillo-Villa, D. E. Soto-Parra, J. A. Matutes-Aquino, R. A. Ochoa-Gamboa, A. Planes, L. Mañosa, D. González-Alonso, M. Stipcich, R. Romero, D. Ríos-Jara, H. Flores-Zúñiga, Phys. Rev. B 83, 174109 (2011).
34. P. O. Castillo-Villa, L. Mañosa, A. Planes, D. E. Soto-Parra, J. L. Sánchez-Llamazares, H. Flores-Zúñiga, C. Frontera, Elastocaloric and magnetocaloric effects in Ni-Mn-Sn(Cu) shape-memory alloy, Journal of Applied Physics 113, 053506 (2013)
35. S. Fahler, U. K. Roßler, O. Kastner, J. Eckert, G. Eggeler, H. Emmerich, P. Entel, S. Muller, E. Quandt, K. Albe, Adv. Eng. Mat., 14, No. 1-2 (2012)
36. F. Kuate Fodouop, G.C. Fouokeng, A. Tsamouo Tsokeng, M. Tchoffo, L.C. Fai, Physica E 128 (2021) 114616
37. M. M. Vopson, Y. K. Fetisov, G. Caruntu, G. Srinivasan, Measurement Techniques of the Magneto-Electric Coupling in Multiferroics, Materials 10, 963 (2017).
38. B. Neese, B. Chu, S-G Lu, Y. Wang, E. Furman, Q.M. Zhang, Large electrocaloric effect in ferroelectric polymers near room temperature, Science **321**, 821-823 (2008).
39. B. G. Shen, J. R. Sun, F. X. Hu, H. W. Zhang, Z. H. Cheng, Recent Progress in Exploring Magnetocaloric Materials, Advanced Materials **21**, Issue 45, 4545–4564
40. B. Nair, T Usui, S. Kurdi, G.G. Guzman-Verri, X. Moya, S. Hirose & N.D. Mathur, Large electrocaloric effects in oxide multilayer capacitors over a wide temperature range, Nature vol 575 2019 468-472.
41. J. Bartlett, G. Hardy & I. D. Hepburn, Performance of a fast response miniature Adiabatic Demagnetisation Refrigerator using single crystal tungsten magnetoresistive heat switch, Cryogenics 72 2017 111-121.